\documentclass[]{aastex631}

\begin{document}

\title{Overview of Hayabusa2 extended mission's flyby of Near-Earth Asteroid (98943) Torifune}

\author[0000-0002-1821-5689]{Masatoshi Hirabayashi}
\affiliation{Georgia Institute of Technology, Atlanta, Georgia, USA}

\author{Masahiko Hayakawa}
\affiliation{Japanese Aerospace Exploration Agency, Kanagawa, Japan}

\author[0000-0002-7181-3522]{Yuya Mimasu}
\affiliation{Japanese Aerospace Exploration Agency, Kanagawa, Japan}

\author{Naru Hirata}
\affiliation{University of Aizu, Fukushima, Japan}

\author[0000-0001-6809-6035]{Takuya Iwaki}
\affiliation{Japanese Aerospace Exploration Agency, Kanagawa, Japan}

\author[0000-0003-1226-9966]{Shunichi Kamata}
\affiliation{Hokkaido University, Hokkaido, Japan}

\author[0000-0002-4809-7492]{Kohei Kitazato}
\affiliation{University of Aizu, Fukushima, Japan}

\author{Toru Kouyama}
\affiliation{National Institute of Advanced Industrial Science and Technology, Tokyo, Japan}

\author[0000-0003-4791-5227]{Naoya Sakatani}
\affiliation{Japanese Aerospace Exploration Agency, Kanagawa, Japan}

\author[0000-0002-4125-0802]{Hajime Yano}
\affiliation{Japanese Aerospace Exploration Agency, Kanagawa, Japan}

\author[0000-0001-6160-9360]{Koki Yumoto}
\affiliation{LIRA, Observatoire de Paris, Université PSL, Sorbonne Université, \\ Université Paris-Cité, CY Cergy Paris Université, CNRS, Meudon, France}

\author{Masahiro Fujiwara}
\affiliation{Japanese Aerospace Exploration Agency, Kanagawa, Japan}

\author{Sumito Shimomura}
\affiliation{Japanese Aerospace Exploration Agency, Kanagawa, Japan}

\author{Takanao Saiki}
\affiliation{Japanese Aerospace Exploration Agency, Kanagawa, Japan}

\author{Hiroshi Takeuchi}
\affiliation{Japanese Aerospace Exploration Agency, Kanagawa, Japan}

\author[0000-0002-6142-9842]{Eri Tatsumi}
\affiliation{Instituto de Astrofísica de Canarias, La Laguna, Tenerife, Spain}
\affiliation{University of La Laguna, Tenerife, Spain}
\affiliation{University of Tokyo, Tokyo, Japan}

\author[0000-0002-9161-6171]{Yuichi Tsuda}
\affiliation{Japanese Aerospace Exploration Agency, Kanagawa, Japan}

\author{Yasuhiro Yokota}
\affiliation{Institute of Science Tokyo, Tokyo, Japan}
\affiliation{Japanese Aerospace Exploration Agency, Kanagawa, Japan}

\author[0000-0002-3118-7475]{Makoto Yoshikawa}
\affiliation{Japanese Aerospace Exploration Agency, Kanagawa, Japan}

\author[0000-0002-4874-0417]{Satoshi Tanaka}
\affiliation{Japanese Aerospace Exploration Agency, Kanagawa, Japan}

\author{Hayabusa2 Extended Mission Torifune Flyby Working Group}



\begin{abstract}
The Hayabusa2 extended mission, nicknamed Hayabusa2\# ($\#$ is pronounced SHARP, which stands for the Small Hazardous Asteroid Reconnaissance Probe), is JAXA's small body explorer to conduct science and engineering investigations in space. After the successful return to the Earth with the samples from the carbonaceous asteroid (162173) Ryugu on December 6, 2020, Hayabusa2 diverted away from Earth to start its decade-long extended mission. The major scope includes engineering demonstration of long-term maintenance strategies for spacecraft and operation systems and scientific investigations during various mission phases. Major scientific investigations include spacecraft-based telescopic observations of exoplanets and zodiacal dust observations during the cruise phase, flyby observations of the near-Earth asteroid (98943) Torifune in July 2026, and rendezvous observations of near-Earth asteroid 1998 KY26 in 2031. This study overviews Hayabusa2\#'s flyby and the physical properties of Torifune. Although the flyby operation planning is still ongoing, the mission will attempt to fly by the target at a distance (from the asteroid’s center) of ~1-10 km. The flyby speed is planned to be 5.25 km/s, while the encounter location is 0.81 au from the sun. {The mission plans to fix the spacecraft's orientation during the flyby, only allowing for a very limited pointing change to attain higher} resolution imaging. The mission will attempt to obtain science and engineering returns during the flyby. The planned investigations will offer stronger insights into material transport mechanisms in the inner solar system and a demonstration of planetary defense technologies. 
\end{abstract}

\keywords{Apollo group (58) --- Asteroids (72) --- Flyby missions (545) --- Near-Earth objects (1092) --- Small Solar System bodies (1469)}


\section{Introduction} \label{sec:intro}

The Hayabusa2 mission was JAXA's second asteroid sample return mission that followed its precursor mission, Hayabusa. The spacecraft was launched in 2014 from JAXA's Tanegashima Space Center and arrived at its nominal mission target, the carbonaceous asteroid (162173) Ryugu, { in 2017} \citep{Tsuda2016, Watanabe2017, Watanabe2024}. While performing detailed remote sensing observations \citep{Watanabe2019, Sugita2019, Kitazato2019, Okada2020}, the mission successfully released the MASCOT rover developed by the German Aerospace Center \citep{Jaumann2019, Mimasu2021} and MINERVA-II-1A and -B \citep{Yoshikawa2022}. It also conducted one small carry-on impact experiment \citep{Arakawa2020} and two sampling operations \citep{Morota2020}. After departing from the asteroid in 2019, the spacecraft spent about one year returning to Earth and delivering Ryugu's samples in 2020. Since then, sample analysis has continuously yielded groundbreaking findings \citep{Yada2022, Nakamura2023, Yokoyama2023}. 

After the remarkable success of the Hayabusa2 mission \citep{Watanabe2019}, later called the nominal mission, the mission restarted as an extended mission nicknamed Hayabusa2\#, where the \# symbol is pronounced like SHARP, which stands for Small Hazardous Asteroid Reconnaissance Probe. At the time of the mission's inauguration, the mission was still in the phase of determining the final scenario \citep{Hirabayashi2021}. The following engineering assessment of the spacecraft revealed that visiting its flyby target, the near-Earth asteroid (98943) Torifune, and rendezvous target, the near-Earth asteroid 1998 KY26, would be a better option \citep{Mimasu2022, Tsuda2025}. 

Selecting the extended mission’s final scenario, which pursues 1998 KY26 as its rendezvous target, took years of effort. The preliminary mission effort identified two mission scenarios for consideration, though detailed engineering assessments remained unresolved at the time \citep{Hirabayashi2021}. The first scenario involved performing a Venus flyby followed by rendezvousing with the near-Earth asteroid 2001 AV43, known as the EVEEA scenario, where E, V, and A stand for Earth, Venus, and Asteroid. In this scenario, the spacecraft would depart from Earth, swing by Venus, and then swing by Earth twice before rendezvousing with its final target. The arrival at 2001 AV43 would be expected to be in late 2029. The other scenario involved performing a flyby at Torifune, followed by a rendezvous with 1998 KY26. This scenario was designated the EAEEA scenario, where the spacecraft would depart from Earth, fly by Torifune, swing by Earth twice, and finally rendezvous with 1998 KY26. The expected timing of its rendezvous with 1998 KY26 is 2031. Later engineering assessments identified constraints on the spacecraft’s trajectory, mainly due to its thermal configuration, which would allow the onboard system to remain operational \citep{Mimasu2022}. The EAEEA scenario was found to be feasible, while the EVEEA scenario would violate thermal constraints due to the spacecraft's original design. The expected trajectory in the EVEEA scenario would reach 0.71 au from the sun, where the spacecraft might be closer to it than Venus. This orbital condition would not satisfy the spacecraft's operations. On the other hand, while the EAEEA scenario would also reach a solar distance of 0.77 au, the trajectory would not violate the constraints. Hayabusa2\# finally identified the EAEEA scenario as its exploration scenario. 

Hayabusa2\# will conduct various observational campaigns in both cruise and critical phases over a decade after its inauguration, while it is defined as an extended mission.  The mission will attempt to advance engineering and scientific innovations for exploring small bodies. The cruise phase is {a} major part of the mission, focusing on various remote sensing observations \citep{Tsumura2023, Yumoto2026, Sano2025} and performance and calibration tests \citep{Yamada2023}. The mission is also dedicated to the development of operational sequences for its flyby and rendezvous phases. Continuous observations of zodiacal light and galactic light \citep{Tsumura2023, Sano2025}, exoplanets \citep{Yumoto2026}, and cometary activity\footnote{\url{https://x.com/haya2_jaxa/status/1833448389310353616}} are key part of the spacecraft's decade-long cruise phase. The mission critical phases include three Earth swing-bys, one of which was completed in 2020, a flyby at Torifune in 2026, and a rendezvous at 1998 KY26 in 2031 (Figure \ref{Fig:Scenario}).

The purpose of this study is to summarize the currently planned flyby operation at Torifune and the objectives for it. Because the Torifune flyby is science-driven, there exist various thrilling, outstanding questions to be discussed. Without detailing such questions, however, this report attempts to make discussions short and provides a concise summary of the flyby plan and its objectives. Outstanding science questions for Torifune are planned to be provided in a separate study. This study also does not detail the whole mission scenario and its science, which is provided in Tanaka et al. (under preparation). This report is organized as follows. Section \ref{Sec:whole_mission} overviews the whole mission sequences until the mission completion. Section \ref{Sec:characteristics} compiles the current understanding of the asteroid's physical characteristics. Section \ref{Sec:objectives} introduces the mission objectives for engineering and science during the Torifune flyby. Section \ref{Sec:flyby_conditions} introduces the planned flyby conditions. Section \ref{sec:remote_sensing} summarizes the onboard remote instruments planned to be used during the flyby operation. Section \ref{Sec:flyby_planning} briefly introduces the current flyby planning. Finally, Section \ref{Sec:planetary_defense} discusses the connection between Hayabusa2\# and Planetary Defense. { Throughout the main texts, we apply the term ``higher resolution" at multiple places to indicate our efforts to increase imaging resolution under highly constraining view geometry during the flyby.}

\section{Hayabusa2\# mission spaceflight overview}
\label{Sec:whole_mission}
\begin{figure}
    \centering
    \includegraphics[width=\linewidth]{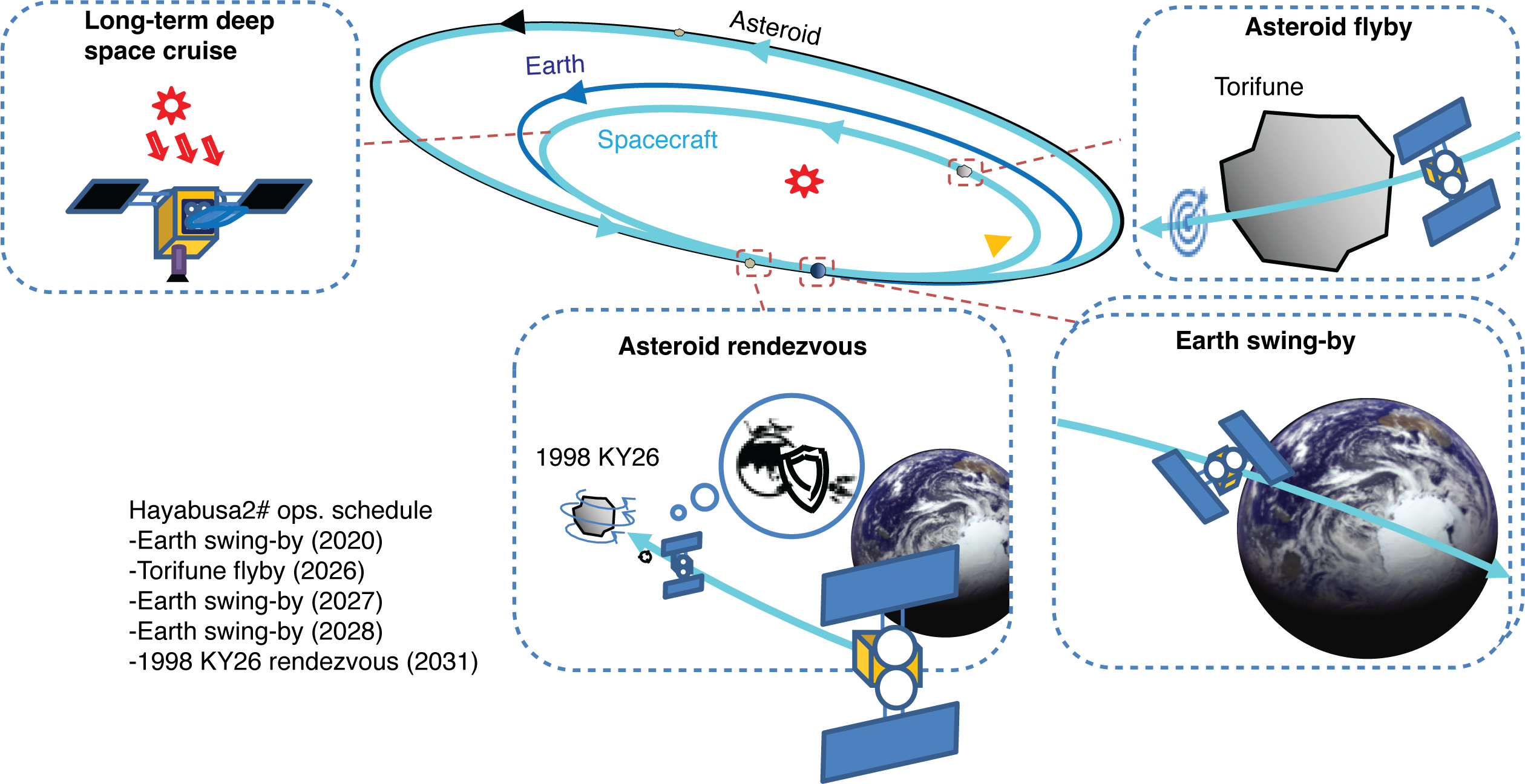}
    \caption{Hayabusa2\# mission spaceflight operation scenario.}
    \label{Fig:Scenario}
\end{figure}

This section briefly summarizes Hayabusa2\#'s spacecraft operation scenario (Figure \ref{Fig:Scenario}). After the successful return of the Hayabusa2 spacecraft with the samples from Ryugu, Hayabusa2\# started its flight operations to complete another decade-long journey. Not only does the extended mission conduct spaceflight operations with science investigations during both cruise and critical phases, but it also supports sample analysis for the returned samples. While sample studies will advance material sciences of primitive materials in the solar system, spaceflight operations will achieve both scientific and engineering exploration by conducting long-term spacecraft operations and by visiting new small bodies. {The mission's spacecraft operations will address classical Planetary Science problems and establish advanced knowledge about Planetary Defense.}

Hayabusa2\#'s spaceflight operations will consist of five key phases (Figure \ref{Fig:Scenario}). The first phase was the first Earth swing-by in 2020, when the nominal mission was completed. The second phase will be the flyby operation at Torifune on July 5, 2026. The third and fourth phases will be Earth swing-bys in 2027 and 2028 to correct the spacecraft's trajectory toward the final destination, 1998 KY26. The final phase will be the rendezvous with 1998 KY26 in 2031. While the overall rendezvous operations are still to be determined, studies proposed possible scenarios \citep{Kikuchi2023, Tsuda2025, Pedros-Faura2025}. Although the details remain undetermined, possible operations without adequate feasibility assessment include determining the gravity field, releasing a target marker, firing a projectile originally for sampling, and landing. A recent study on the mission also identified a possible trajectory reaching the near-Earth asteroid 2024 YR4 with limited trajectory corrections. However, at present, the current decision on the mission is to keep and complete its original plan (Nakagawa et al., under review). {As a side note, 2024 YR4's collision with the Moon was reported to be ruled out \citep{Rivkin2026}.} 

\section{Torifune physical characteristics}
\label{Sec:characteristics}

After officially approved by the International Astronomical Union (IAU), 2001 CC21 (the former designation) was named (98943) Torifune. This asteroid was discovered by the Lincoln Near-Earth Asteroid Research (LINEAR) project. The name originated from god in Japanese mythology. It also means the god's ship, which travels safely at high speed, like a bird, and is as steady as a rock.\footnote{\url{https://www.hayabusa2.jaxa.jp/en/topics/20240925_2001_CC21_e}} This name was selected with the support of the Children's Selection Committee in Japan, which included nine members ranging from elementary students to junior high school students. 

Since Hayabusa2\# identified Torifune as its flyby target candidate, the asteroid has been actively investigated through ground- and space-based telescopic observations. Depending on telescopic techniques, measurements are found to fluctuate and offer different uncertainties. Torifune is a near-Earth asteroid (NEA) with an orbit similar to that of the Apollo-type group. The definition of the Apollo group is that the orbital semi-major axis is higher than 1 au, while the periapsis is shorter than 1 au.\footnote{\url{https://www.minorplanetcenter.net/iau/Unusual.html}} According to the JPL Horizons System,\footnote{\url{https://ssd.jpl.nasa.gov/horizons/app.html}} Torifune's orbital semi-major axis is 1.032 au, and its periapsis is 0.806 au. Furthermore, the inclination is 4.81$^\circ$, and the orbital period is about 383 days. Yarkovsky's $A_2$ parameter is also available as $1.35 \times 10^{-14}$ au/day$^2$. The current position uncertainty at the time of the planned flyby is 14.6 km. All the quantities provided are based on the solution at 08:09:18 on October 16, 2025. 

The preponderance of evidence suggests that the asteroid's equivalent diameter (the diameter of a sphere with an equivalent volume) is $\sim450$ m, while the shape may be elongated. Analysis using occultation data on March 5, 2023, suggests the asteroid has a semi-major axis of $420^{+80}_{-60}$ km and an aspect ratio of $b/a = 0.37^{+0.09}_{-0.09}$, where $a$ is the semi-major axis and $b$ is the semi-minor axis \citep{Arimatsu2024}. An observational study using Spitzer and ground-based observations determined an equivalent diameter of $465_{-15}^{+15}$ m and an $b/a$ value of 0.57 \citep{Fornasier2024}. \cite{BourdelledeMicas2025} performed photometric observations to measure an equivalent diameter of $465^{+15}_{-15}$ m. Another study by light curve inversion suggested an equivalent diameter of $440^{+60}_{-60}$ m and an ellipsoidal dimension of $840^{+160}_{-120} \times 320^{+100}_{-80} \times 340^{+60}_{-60}$ km$^3$ \citep{Popescu2025}. Furthermore, \cite{Fatka2025} suggested that the aspect ratio of this asteroid would be $b/a = 0.63$ and $c/a = 0.47$. A study using the Mid-Infrared Spectrograph and Imager (MIRSI) at the 3-m NASA Infrared Telescope Facility (IRTF) employed optical and thermal measurements to determine an equivalent diameter of $433_{-34}^{+36}$ m \citep{Lopez-Oquendo2025}. It also utilized the optical parameters published on the JPL Horizons to obtain $398_{-9}^{+11}$ m \citep{Lopez-Oquendo2025}. NEOWISE observations from November 2021 through February 2024 suggest that an equivalent diameter of 337$^{+33}_{-27}$ m \citep{Wright2025}.

The reported geometric albedo ranges between 0.2 and 0.35. \cite{Geem2023} reported that the albedo would be $23^{+4}_{-4}$\% estimated by polarimetric approach. Furthermore, \cite{Fornasier2024} showed that it would be $21.6^{+1.6}_{-1.6}$\%. The analysis by \cite{Lopez-Oquendo2025}, using optical and thermal measurements derived from data acquired on January 22, 2023, led to an albedo of $25_{-9}^{+9}$\% (L{'}opez-Oquendo, personal communication). They also pointed out that using the optical parameters from the JPL Horizons System led to an albedo of $30^{+15}_{-12}$\%. \cite{Lopez-Oquendo2025} further offered other albedo solutions, based on data acquired later days, such as February 7, 2023 and February 2, 2024, up to $64$\%. However, the object was fainter during these observations, so the outcomes might not be reliable. They further pointed out that the errors in absolute magnitude might be the major source of albedo uncertainties \citep{Masiero2021}, also causing the errors in diameter measurements and vice versa. 

Light curve studies constrained the spin period tightly, although the spin orientation solutions show some variations. \cite{Fornasier2024} reported $5.02124 \pm 0.00001$ hours. The measured spin period by \cite{Popescu2025} is $5.021516 \pm 0.000106$ hours. Their spin pole solution is $\beta = {+89_{-6}^{+1}}^\circ$ and $\lambda = {301^{+35}_{-35}}^\circ$, where $\lambda$ and $\beta$ are the ecliptic longitude and latitude. On the other hand, the spin period provided by \cite{Fatka2025} is $5.021522 \pm 0.000003$ hours. The spin axis orientation is found to be $\beta=+84^\circ$ and $\lambda = 259^\circ$. In the reported representation, the uncertainties of the spin axis orientation show a larger separation from the mean values. The consistency of the spin orientation is that it likely faces the ecliptic north. At this point, there is no report giving tumbling mode detection. In addition to light curve observations, an analysis using NEOWISE data also prefers prograde rotation with an obliquity of ${24^{+6}_{-9}}^{\circ}$.

The understanding of Torifune's taxonomy has expanded since it was identified as Hayabusa2\#'s target. The earlier taxonomic class was reported to be an L-type \citep{Binzel2004}. Early mission analysis applied this classification \citep{Hirabayashi2021}. If this is true, Torifune might have a high content of the FeO-bearing spinel mineral, which is related to {Calcium-Aluminum-rich Inclusions (CAIs)}, interpreted as the oldest mineral in the solar system. However, other spectral observations suggested the asteroid is S-complex, characterized by clear absorption features at 0.9-1 µm and weak ones at 2 µm \citep{Lazzarin2005}, related to the presence of pyroxene and olivine. In \cite{DeMeo2009}, the asteroid was classified as an Sw-type in the Bus-DeMeo classification, addressing its higher slope than a typical S-type slope. Spectroscopic and spectrophotometric investigations favor this asteroid's taxonomy as S-complex, likely an Sq-type \citep{Geem2023}. Based on the mineralogy of reflectance spectra, the asteroid may be related to either L- or LL-type chondrites \citep{Geem2023, Fornasier2024, Popescu2025}, while a study favored its connection with an L-chondrite \citep{BourdelledeMicas2025}. Therefore, Torifune appears to have similar compositions to as asteroids Itokawa and Eros, as well as Dinkinesh \citep{Levison2024}, which are all members of the S-type taxonomic class and have affinities to L- and LL-type chondrite meteorites.

{
\section{Planetary protection}

With the detailed telescopic characterization, Hayabusa2\# proposed a categorization policy for Torifune, as well as that for 1998 KY26 \citep{Yano2024}. Planetary missions are expected to comply with the COSPAR Policy of Planetary Protection (PPP) \citep{COSPAR-PPP2024}. In contrast to the Ryugu case during the nominal mission, the extended mission needs to resolve the case that the spacecraft has already been flying to reach out to its targets, making it impossible to develop additional sterilization countermeasures at the time of PPP consideration. The mission identified that Category I would be appropriate for the Torifune flyby. In this category, a proposed mission is supposed to be not of direct interest for understanding the process of chemical evolution and/or the origin of life; therefore, no protection of a target body is thus warranted, and no planetary protection guidelines are imposed by this policy \citep{COSPAR-PPP2024}. As a note, the mission considered Category II would be appropriate for 1998 KY26, in which a mission may identify significant interest relative to the process of chemical evolution and/or the origin of life, but scientific opinion provides a remote chance of contamination by organic or biological materials which could compromise future investigations of the process of chemical evolution and/or the origin of life \citep{COSPAR-PPP2024}. The COSPAR Panel on Planetary Protection approved this request on May 12, 2025. The planned investigations on the mission comply with the policy aligned with the approved category.}

\section{Mission objectives for Torifune flyby investigations}
\label{Sec:objectives}

The spacecraft is not designed for fast flyby operations (it was originally designed for rendezvous operations) and is not capable of fast slews to keep the asteroid in its field-of-view (FOV) during closest approach. This unique encounter is pivotal for the operations team to examine how it can achieve the mission's goals and maximize science and engineering returns during the flyby under such constraints. One note is that, because Hayabusa2\# is an extended mission, it applies a bottom-up approach. In other words, it does not prioritize planned operations over the mission lifespan but continuously accumulates successes from earlier phases. 

\subsection{Engineering objectives}

The engineering objectives stem from its long-term mission ($>$10 years) after the completion of the nominal mission: (1) exploring how to resolve issues related to the system's aging; (2) establishing fast flyby technologies; and (3) extending the rendezvous experience to an extremely small fast rotator. The second engineering objective, establishing fast flyby technologies, is aligned with the Torifune flyby investigation and critical in terms of using a spacecraft system that was not designed for flyby operations. As stated before, it lacks adequate capability to slew during the closest approach, and not all onboard instruments are aligned with the preferred directions during the approach phase. The recent flyby planning favors a close encounter that possibly reaches less than a few km from the asteroid's center; the current reference encounter distance is 1-10 km to compensate for the spacecraft's slew rate limitation, but it can be closer depending on the ongoing flyby planning. Achieving the planned close encounter requires accurate orbit determination for Torifune prior to the arrival, which needs tight collaboration with astronomers who have provided telescopic observations, and robust spacecraft guidance, navigation, and control (GNC). The mission will offer a demonstration opportunity for flyby operations of an aging space probe that was not originally designed for such operations. Given the planned flyby condition, in which the spacecraft will reach a distance of 1-10 km from the target center, the mission anticipates an additional challenge compared to the kinetic impactor scenario. The spacecraft stays on course to hit the target until very close to the collision, but needs to be off course and pass by the target at a short distance safely. The mission identifies the planned flyby sequences as a demonstration of technology related to not only kinetic deflection, i.e., sending a space probe to make a collision with a target and push it from the course to hit the Earth \citep{Tsuda2025}, but also the fast reconnaissance concept, which is a new concept recently addressed in Planetary Defense, which is detailed in Section \ref{Sec:planetary_defense}.

\subsection{Scientific objectives}
\label{Sec:sci_obj}
The scientific objective for the Torifune flyby observations is to identify the asteroid's physical and geological properties under limited observational conditions. Because of the expected high-speed flyby, the mission needs to identify constraints on satisfying all the instruments' observational conditions. While onboard instruments having {larger} FOVs can observe the target most of the time prior to the encounter, some instruments may {be able to obtain limited amounts of data due to the nature of the flyby}. To reflect this observational constraint, the mission emphasizes its attempt to allow all onboard instruments to obtain at least one data point during the encounter. Hayabusa2\# will explore the following two science questions: material transport mechanism in the inner solar system; and establishment of advanced knowledge and technologies for Planetary Defense \citep{Hirabayashi2021}. Because Planetary Defense is interdisciplinary, science and engineering efforts on the mission can better emphasize this issue. The mission's Planetary Defense effort is also aligned with the recent reports addressing the importance of the rapid reconnaissance concept, a recently identified coordination effort that aims to determine the physical properties of a potentially hazardous object when it is discovered to threaten the Earth within a limited time \citep{Abell2020, Barbee2020, NAS2022}.

Table \ref{tab:science_obj} summaries the mission's science objectives for the Torifune flyby. The mission plans to employ all but one onboard instrument (Section \ref{sec:remote_sensing}), and each science objective will be achieved given the contribution of each designated instrument. The science objectives encompass investigations that detail geophysical, geological, thermal, and geochemical characterizations of Torifune. Observing Torifune's detailed morphological features, such as textures, mass wasting, and possible fractures, at {higher} resolution will also offer crucial constraints on its interior, providing pivotal insights into the formation and evolution processes of subkilometer NEAs. Furthermore, the planned flyby will offer strong synergies between telescopic and flight observations. Comparative studies between telescopic and flight observations will provide stronger insight into the target asteroid and also validate both aspects of the observations. 

The science objectives in Table \ref{tab:science_obj} have evolved since the start of the extended mission in 2020. The recent advances in understanding this asteroid through telescopic observations have enabled the mission to update its views of Torifune. As mentioned above, Torifune was assumed to be an L-type asteroid during the mission's design phase; the original geochemical investigations aimed to explore an L-type asteroid. However, since it was selected as the mission's flyby target, telescopic observations have provided insight into the asteroid's taxonomic class, which favors an S-type over an L-type (Section \ref{Sec:characteristics}). The mission does not rule out the L-type taxonomy because the overall composition is still undetermined and L-type materials could be distributed locally on the asteroid.

\begin{table}[]
    \centering
    \caption{Science objectives (SOs) for the Torifune flyby. The SOs are not listed in priority order. ONC stands for the Optical Navigation Camera. It includes both a telescopic camera (ONC-T) and a wide-angle camera (W1), but the ONC-T is the main contributor to the imaging listed SOs. TIR is the thermal infrared imager. NIRS3 is the Near-Infrared Spectrometer. Finally, LIDAR is the laser altimeter.}
    \begin{tabular}{lll}
       \hline
       SO ID & Instrument/Task & Description \\ 
       \hline
       \hline
       SO.~1 & ONC & Perform imaging observations to reveal the spin axis state and reflectance \\
       SO.~2 & ONC & Perform imaging observations to capture the asteroid's global surface features \\
       SO.~3 & TIR & Perform thermal imaging observations to determine the thermal properties \\
       SO.~4 & NIRS3 & Conduct spectral observations to determine surface compositions \\
       SO.~5 & LIDAR & Attempt to obtain sampling laser altimeter data \\
       SO.~6 & Shape modeling & Reconstruct the shape using flyby data \\
       \hline
    \end{tabular}
    \label{tab:science_obj}
\end{table}

\section{Torifune flyby conditions}
\label{Sec:flyby_conditions}
\begin{figure}[ht!]
    \centering
    \includegraphics[width=\linewidth]{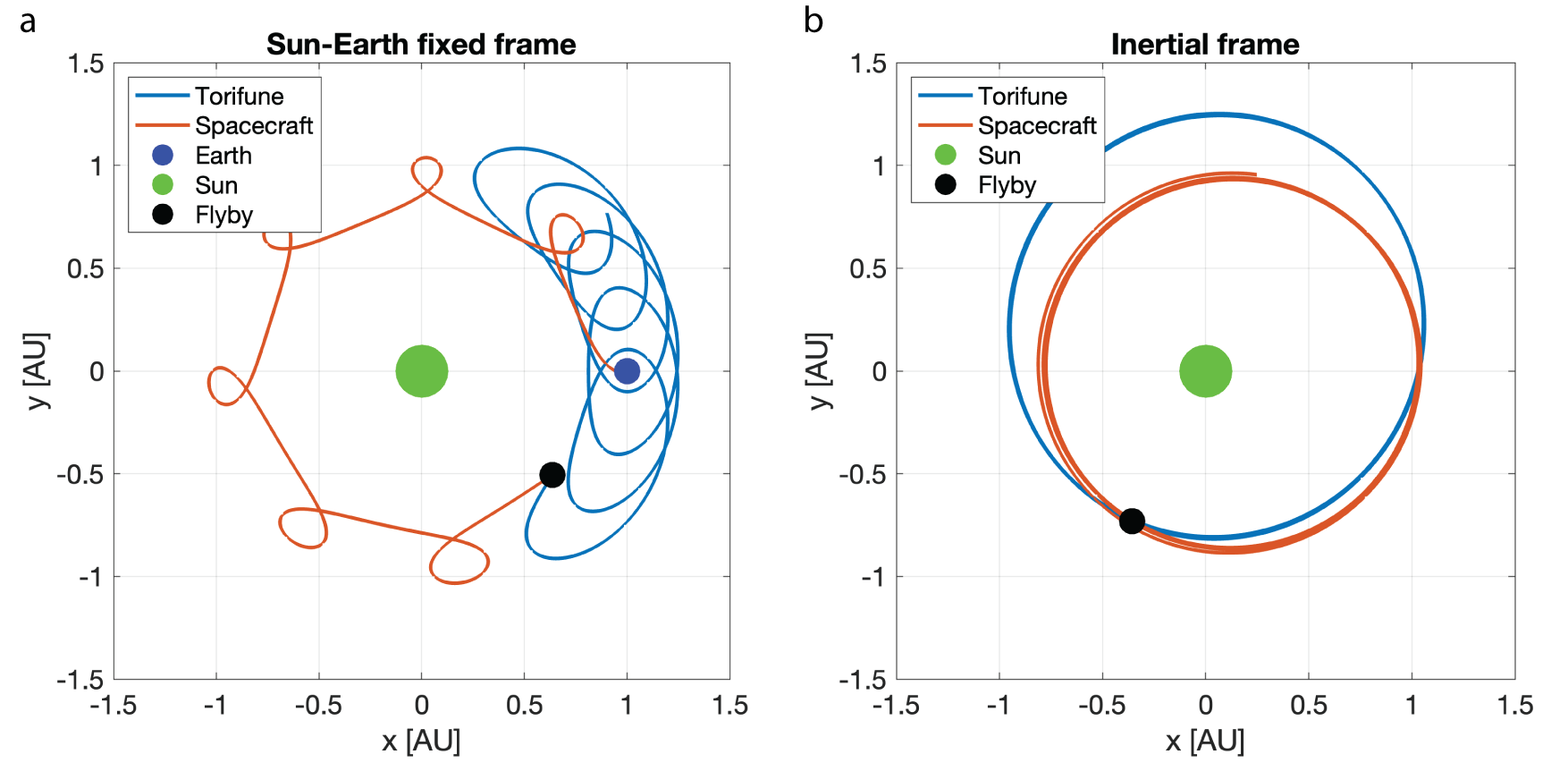}
    \caption{Spacecraft trajectory to Torifune from Earth departure in December 2020 to the flyby in July 2026. The spacecraft's trajectory is shown as red lines, while Torifune's trajectory is depicted using blue lines. The blue {dot} is Earth, the green {dot} is the sun, and the black {dot} is the flyby point. Both panels show the same trajectories. Panel a shows the inertial frame, while Panel b illustrates the sun-Earth fixed frame.}
    \label{Fig:trajectory}
\end{figure}

Since the return to Earth with Ryugu's samples, the Hayabusa2\# spacecraft has conducted multiple Ion Engine System (IES) operations. The IES will continue to gain the necessary delta-V to reach Torifune until May 2026. The currently planned arrival is scheduled for July 5, 2026. Because the solar distance at the time of the spacecraft flyby is $\sim$0.814 AU, the spacecraft will experience higher thermal severity than the original design. However, detailed assessments during the mission's target selection revealed that the spacecraft's thermal capability would satisfy the flyby condition \citep{Mimasu2022}. {The spacecraft encountered minor issues with the IES system and its reaction wheels; however, the mission has identified resolutions, and no risks are seen prior to the execution of the flyby.} Figure \ref{Fig:trajectory} shows the trajectories of the spacecraft and asteroid in two coordinate frames. 

\begin{figure}[ht!]
    \centering
    \includegraphics[width=0.5\linewidth]{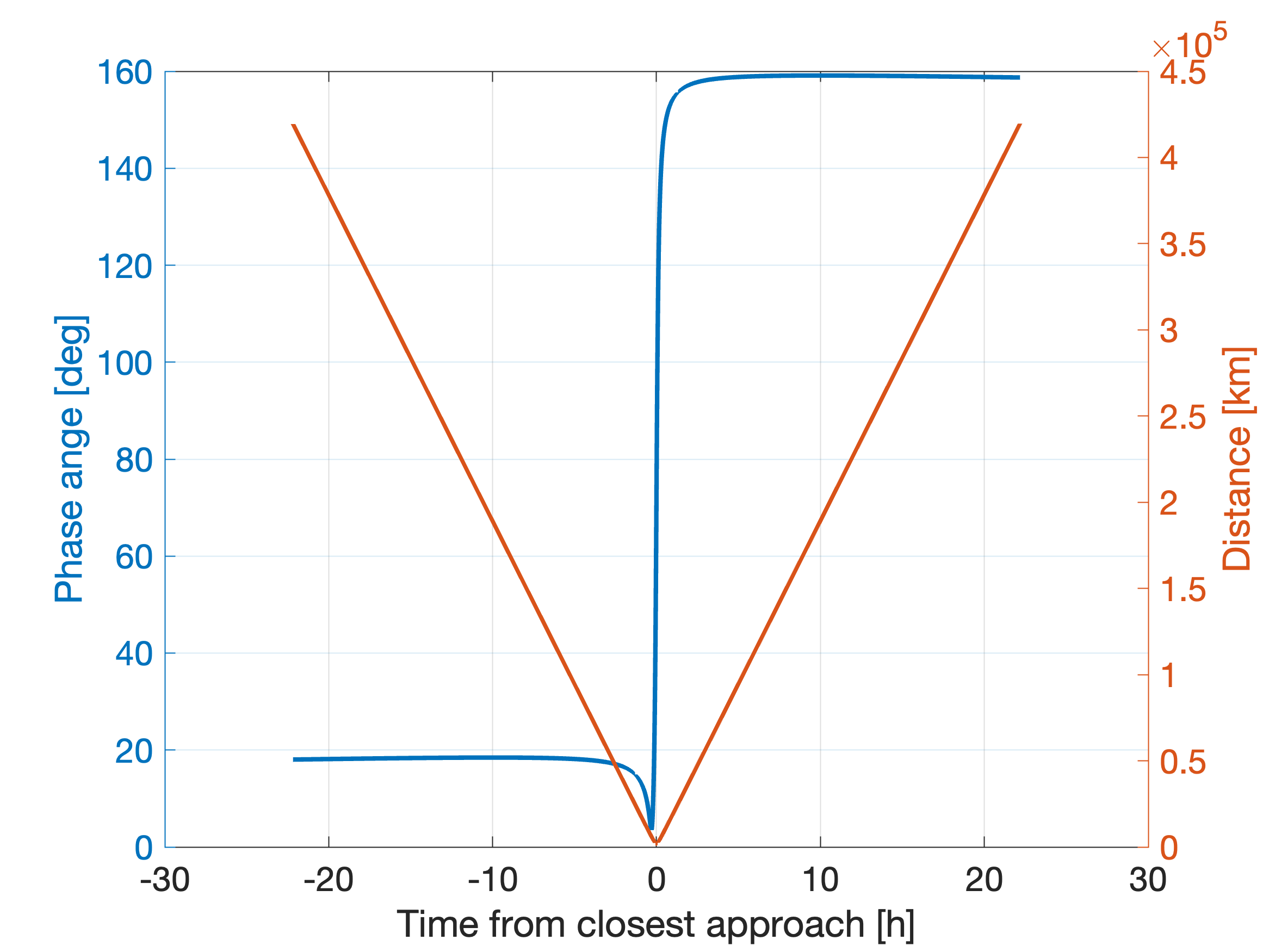}
    \caption{Variations in the solar phase angle (spacecraft-asteroid-sun angle) and the spacecraft-asteroid distance over $T \pm 30$ hr from the closest encounter.}
    \label{fig:phase_angle}
\end{figure}

Just before the flyby, the spacecraft will orbit inside the asteroid's heliocentric orbit (Figure \ref{Fig:trajectory}). It will then approach Torifune from the inside toward the outside (of the asteriod's orbit). The flyby condition is listed in Table \ref{Tab:Flyby_conf}. The encounter speed is planned to be 5.25 km s$^{-1}$. Because of this relative trajectory, Torifune's sunlit region will face the spacecraft before the closest approach, but its shadowed region will be dominant in the {spacecraft's viewing geometry} after the encounter. The expected solar phase angle (spacecraft-asteroid-sun) will be $\sim20^\circ$ until 5 hr before the closest approach (T-5 hr). The solar phase will then gradually decrease while the spacecraft approaches the asteroid after the closest approach, the phase angle will immediately reach up to $\sim160^\circ$ (Figure \ref{fig:phase_angle}). This viewing geometry, therefore, only allows the spacecraft to observe Torifune before the encounter, essentially prohibiting post-encounter spacecraft observations. 

The planned flyby sequences will begin approximately T-10 days after the spacecraft successfully detects Torifune {(Figure \ref{fig:science_seq})}. The sequences will consist of three phases. The first phase will be the ground-based GNC phase. This phase will continue from T-10 days until T-12 hr. The spacecraft will perform trajectory correction maneuvers (TCMs) by actively employing ground-based and optical navigation. In this phase, therefore, the ground station will be on the loop to help the spacecraft navigate itself toward the asteroid. The ground-based and onboard GNC phases came from the nominal mission's experience during touchdown operations on Ryugu \citep{Terui2022}. The second phase will be the onboard GNC phase. This phase will continue from T-12 hr until T-5 min, where the spacecraft will focus on TCMs based on optical relative navigation. The spacecraft's operations will be performed onboard only. The final phase will be the encounter phase. It will be between T-5 min and the closest encounter. During this phase, the spacecraft will focus on science operations. Again, the closest approach distance and orientation are still to be determined. However, the reference flyby distance from the center of the asteroid is 1-10 km. Conducting a flyby in this distance range will achieve one of the closest flybys targeting small bodies at high speeds ($>$ 1 km s$^{-1}$) made by planetary missions so far (Table \ref{Tab:closestdist}).

\begin{table}[]
    \centering
    \caption{Torifune Flyby tentatively planned conditions}
    \begin{tabular}{lll}
        \hline
        Condition & Description & Quantity \\
        \hline
        \hline
        Flyby timing & Month/year of the planned flyby & July 5, 2026 \\
        Flyby speed & Speed at the closest approach & 5.25 km s$^{-1}$ \\
        Flyby distance & Closest distance & $1-10$ km \\
        Solar distance & Distance from the sun & 0.81 AU \\
        Pre-flyby solar phase & Phase angle 10 h before flyby & 19$^\circ$ \\
        Post-flyby solar phase & Phase angle 10 h after flyby & 158$^\circ$ \\ 
        \hline
    \end{tabular}
    \label{Tab:Flyby_conf}
\end{table}

\begin{table}[]
    \centering
    \caption{List of flyby distances performed by spaceflight missions. The listed distances are for small body flybys at high speeds, which are beyond 1 km s$^{-1}$.}
    \begin{tabular}{llll}
        \hline
        Mission & Target & Distance [km] & Reference \\
        \hline
        \hline
        {Lucy} & {(52246) Donaldjohanson} & {961.4 km} & {Marchi et al. (under review)} \\
        {Hera} & {Deimos} & {$\sim$1000 km} & {ESA website\footnote{\url{https://www.esa.int/Space_Safety/Hera/Hera_asteroid_mission_spies_Mars_s_Deimos_moon}}} \\
        Lucy & (152830) Dinkinesh & 431 km & \cite{Levison2024} \\
        New Horizons & (486958) Arrokoth & 3,538 km & \cite{Stern2019} \\
        Chang’E-2 & (4179) Toutatis & $0.77 \pm 0.12$ km (from surface) & \cite{Zou2014} \\ 
        EPOXI & 103P/Hartley 2 & 694 & \cite{Snodgrass2022} \\
        Deep Impact & 9P/Tempel 1 & 575 & \cite{Snodgrass2022} \\ 
        Rosetta & (21) Lutetia & 3,170 & \cite{Schulz2012} \\
        Rosetta & (2867) Steins & 802 & \cite{Leyrat2010} \\
        Stardust-NExT & 9P/Tempel 1 & 181 & \cite{Snodgrass2022} \\
        Stardust & 81P/Wild 2 & 237 & \cite{Snodgrass2022} \\
        Stardust & (5535) Annefrank & 3,078.5 & \cite{Newburn2003} \\ 
        Deep Space 1 & 19P/Borrelly & 2,171 & \cite{Snodgrass2022} \\ 
        Deep Space 1 & (9969) Braille & 26 & \cite{Oberst2001} \\ 
        NEAR & (253) Mathilde & 1,212 & \cite{Veverka1997} \\
        Galileo & (243) Ida & 2391.2 & \cite{Belton1996} \\
        Galileo & (951) Gaspra & 1,600 & \cite{Veverka1994} \\
        Suisei & 1P/Halley & 152,400 & \cite{Snodgrass2022} \\
        Giotto Extended Mission & 26P/Grigg-Skjellerup & $<200$ & \cite{Snodgrass2022} \\
        Giotto & 1P/Halley & 605 & \cite{Snodgrass2022} \\
        Sakigake & 1P/Halley & $7 \times 10^6$ & \cite{Snodgrass2022} \\
        Vega 2 & 1P/Halley & 8,030 & \cite{Snodgrass2022} \\
        Vega 1 & 1P/Halley & 8,889 & \cite{Snodgrass2022} \\
        ICE & 21P/Giacobini-Zinner & 7,862 & \cite{Snodgrass2022} \\
        \hline
    \end{tabular}
    \label{Tab:closestdist}
\end{table}

\section{Remote sensing instruments and measurement plans}
\label{sec:remote_sensing}

The spacecraft has four remote sensing instrument systems: the Laser Altimeter (LIDAR) \citep{Mizuno2017, Yamada2017}, the Near-Infrared Spectrometer (NIRS3) \citep{Iwata2017}, the optical navigation camera (ONC) system \citep{Kameda2015, Kameda2017, Suzuki2018}, and the thermal infrared imager (TIR) \citep{Okada2017, Arai2017}. The ONC system comprises three framing cameras: ONC-T, ONC-W1, and ONC-W2. The ONC-T is a telescopic camera, while ONC-W1 and ONC-W2 are wide-angle cameras. Table \ref{Tab:remote_instruments} summarizes the properties of these instruments. {These instruments have shown degradation mainly due to the touchdown operations on Ryugu. ONC's sensitivity decreases to $\sim$90\% of the launch level \citep{Yamada2023}, while TIR experiences higher background digital numbers (DNs) by $\sim$20 DN, almost equivalent to a $\sim$2-K background temperature increase for a 300-K black body. NIRS3's and LIDAR's sensitivity decreased by $\sim$23\% and $\sim$10\% after the second touchdown, compared to their initial performance. However, none of the instruments has shown a significant change in performance since then. Thus, the mission expects that appropriate correction procedures can mitigate the reported degradation.} The mission plans to conduct observations using five onboard remote sensing instruments during the Torifune flyby: ONC-T, ONC-W1, TIR, NIRS3, and LIDAR. The mission has decided not to use ONC-W2, given that its viewing geometry was not favored to capture Torifune. For the scientific purposes, each instrument defines its measurement plans that are aligned with the defined science objectives.

The ONC-T measurement plans will consist of four telescopic observations: 1. Light curve observations at a signal-to-noise ratio (S/N) of $>$50, in contrast to ground observations at S/N$\sim$20; 2. Multi-band photometric observations within a wavelength between 400 and 950 nm at S/N$>$200 to see longitudinal variations in surface compositions; 3. Satellite search, in which confirming the existence of a satellite/satellites enables system mass estimation; 4. Geologic and shape observations through disk-resolved imaging during the encounter phase. The first and fourth observation plans will address SO.1, while the second through fourth observation plans will be related to SO.2. The instrument's observation will start before the ground-based GNC phase (T-14 days) until the closest encounter. In the period from T-4 days to T-15 min, during which Torifine is detected as a 1-pix light source, the instrument will perform light curve observations. During the encounter phase, ONC-T will explore the geological characteristics of this asteroid. It will also attempt to determine the spin axis and and surface reflectance. ONC-W1 will contribute to the satellite search.

TIR's measurement plans include determining the thermal properties, including the thermal inertia and surface roughness \citep{Okada2020, Rozitis2020}. { Thermal inertia determines how easily a target can be heated/cooled. On the other hand, surface roughness controls thermal emission. These quantities correlate with each other but can be decomposed using detailed analysis. Another question is whether thermal inertia depends on asteroid size. Recent work compiled a correlation between thermal inertia with asteroid size, though samples were larger than km \citep{MacLennan2021}. Successful thermal imaging by TIR will add a new data point for this question.} Capturing both sunlit and shadow regions can offer temperature variations, giving tighter constraints on the thermal properties of surface materials. The instrument will be used to obtain the thermal fluxes during both nighttime and daytime, which constrains the asteroid's rotational phase at the time of the flyby. Our current plan is for the spacecraft to fly by the target when the morning side with some shadowed regions is visible. 

NIRS3's measurement plans include receiving spectroscopic data during the Torifune flyby. The mission will set the spacecraft's orientation favorable for this instrument to maintain its observations. During the flyby, NIRS3 will maintain an S/N higher than 10, which is the defined minimum threshold for data quality. This instrument will perform both disk-integrated observations before T-50 sec and disk-resolved observations after that time. Data sampling will be performed every 0.64 s. With the obtained data, NIRS3 will analyze the existence of a 3-$\mu$m absorption signature, which is related to the existence of water and hydroxyl, H$_2$O/OH, to infer how the asteroid's surface has been influenced by solar wind-driven water. NIRS3 will also determine the compositional characteristics of Torifune. It can also determine whether Torifune's taxonomy is an S-complex or L-type. If Torifune is S-complex, it may primarily consist of L- or LL-chondrites. While the L-type taxonomy may be not likely, it may host the related materials, like CAI, locally. NIRS3 may be able to identify such compositional signatures. 

LIDAR's measurement plans include obtaining at least one pulse reflected from the asteroid. Given the instrument's detection limit up to 25 km via its far-system mode for the C-type asteroid case, the expected time span that LIDAR can receive reflected pulses is just a few seconds. {Given the suggested taxonomy for Torifune, which may give a higher geometric albedo, this 25-km limit is a conservative value.} During its operations, LIDAR will emit a laser pulse every 1 sec. The spacecraft's orientation after T-5 min until the completion of the flyby will account for LIDAR's viewing condition so that the instrument can capture the asteroid. Because the mission expects only a few data points at maximum, how such data points can offer scientific and engineering contributions is still to be determined. However, obtaining the accurate altitude at the time of observation may help other instruments determine their observational conditions. Another possibility is to be able to characterize the albedo features at the 1,064 nm wavelength. In addition, the LIDAR data may contribute to the orbit position accuracy of Torifune. However, the current view is that, given limited data points, this process may not improve the accuracy much. Applying the concept of the fast reconnaissance concept (Section \ref{Sec:planetary_defense}), the mission considers that making an effort to observe Torifune using LIDAR is our demonstration of its use under limited conditions. 

While this is not from direct observations, shape modeling is key part of the mission's primary investigation plan. Shape modeling will be performed using data from multiple instruments. The major data sets will be ONC-T imagery, while TIR imagery will also be useful. If samples from LIDAR are available, the process may be able to determine the target's size accurately. The drawback of shape modeling during the flyby is that the planned closest flyby distance will not change the viewing geometry from the spacecraft, preventing a better constraint on the topographic information along the line-of-sight. This issue on Hayabusa2\# will be more severe than that on the recent flyby operations by New Horizons at (486958) Arrokoth \citep{Stern2019} and by NASA's Lucy at (152830) Dinkinesh \citep{Levison2024}. The mission plans to use multiple data sets to attempt a better shape reconstruction. The team seeks various approaches, such as light curve inversion, shape-from-silhouette (limb profiling), structure-from-motion, and stereophotoclinometry, to achieve this plan.

\begin{table}[]
    \centering
    \caption{Instrument geometric properties. FOV stands for the field of view. The FOVs of ONC-W1 and W2 and TIR are listed in order. LIDAR's FOVs are listed for long-distance use (left) and short-distance use (right). ONC-W2 will not be used during the flyby.}
    \begin{tabular}{lllll}
        \hline
        Instrument & FOV [$^\circ$] & Pixel size [pix] & Notes \\
        \hline
        \hline 
        ONC-T      &  6.27 & $1,024 \times 1,024$ & Telescopic camera, seven bandpass filters\\
        & & & and a panchromatic window \\
        ONC-W1, W2 & 69.71, 68.89 & $1,024 \times 1,024$ & Wide-angled cameras \\ 
        TIR        & $16.6 \times 12.7$ & $328 \times 248$ & Thermal imager \\
        NIRS3      & 0.11 & 1 & Point-source spectrometer, 128 channels \\
        & & & with 18 nm per pixel \\ 
        LIDAR      & 0.086, 1.15 & 1 & Laser altimeter (a YAG laser at 1,064 nm) \\
        \hline
    \end{tabular}
    \label{Tab:remote_instruments}
\end{table}

\section{Planned flyby science observation sequences}
\label{Sec:flyby_planning}
\begin{figure}[ht!]
    \centering
    \includegraphics[width=0.8\linewidth]{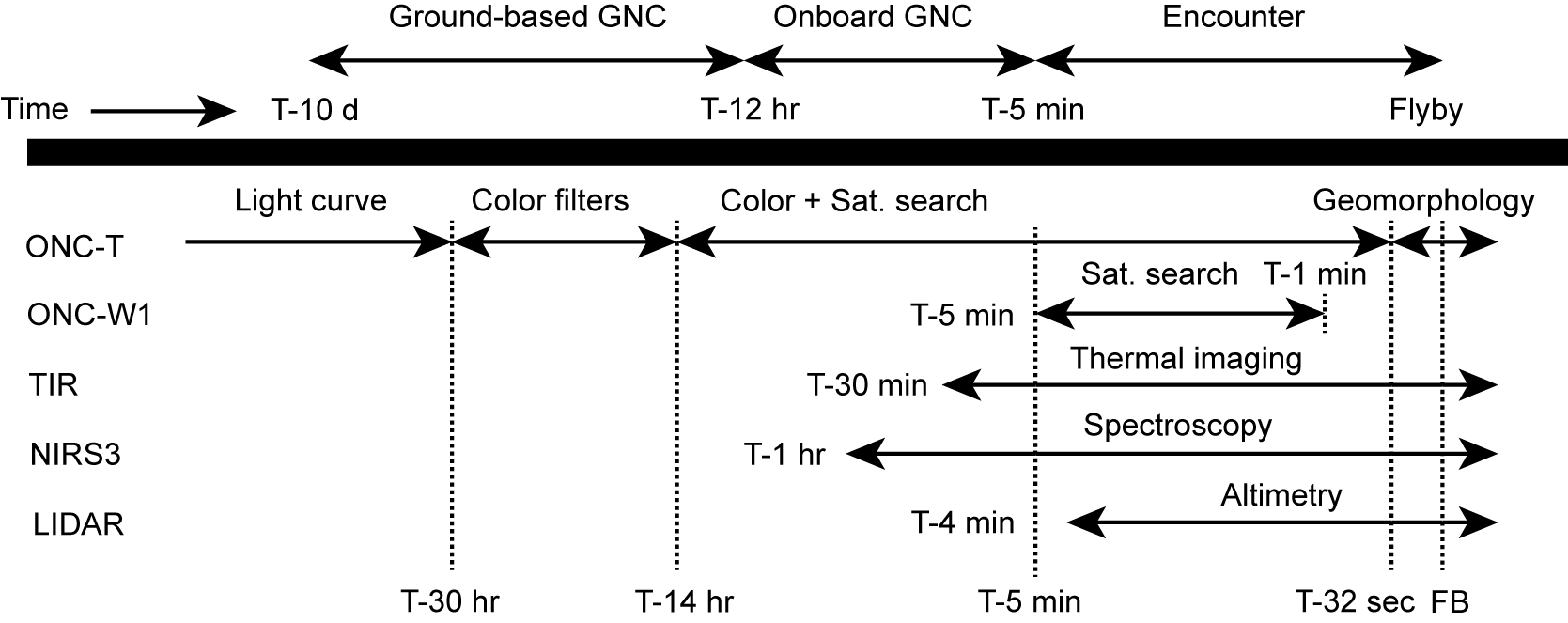}
    \caption{Tentative planning for flyby science observation sequences.``Sat. search" means satellite search. The plot displays time on the horizontal axis, with each row representing the timeline of scientific observation sequences for each instrument. ONC-T will be used not only for scientific investigations but also for GNC sequences. FB means the closest-approach time. All the instruments will complete their observations from 0 sec to 10 min after the closest approach because the spacecraft can no longer observe Torifune.}
    \label{fig:science_seq}
\end{figure}

As the flyby opportunity approaches, the mission has been developing the Torifune flyby sequences that comply both engineering and science sequences. As of the time of this study, the planning effort is still ongoing. While the details are to be determined and will be reported in a separate study, this section summarizes the current status, focusing on scientific investigation. As stated earlier, the flyby sequence is split into three phases (Figure \ref{fig:science_seq}). The phase between T-10 days and T-12 hr is the ground-based GNC phase. The onboard GNC phase is between T-12 h and T-5 min. The final phase is the encounter phase, spanning from T-5 min until the closest flyby. The mission plans to employ ONC (T, W1), TIR, NIRS3, LIDAR to fulfill the science objectives. Again, the mission has decided not to use ONC-W2, given its viewing condition not favored to observe Torifune. Figure \ref{fig:science_seq} illustrates a brief overview of the planned science observation sequences for the onboard remote sensing instruments. The current plan defines the encounter phase as a science-focused phase, but also accepts scientific sequences beyond this phase. The instruments will be operational earlier, so they can be ready for observations after T-5 min or even start observations much earlier than that if slots are available.

ONC-T will be operational continuously during the flyby, not only for scientific observations but also for the spacecraft's GNC sequences. It will conduct light curve measurements until the middle of the ground-based GNC phase. After the light curve measurements, ONC-T will start observing the targets with multiple color filters. The instrument will observe the asteroid at various rotational phases using the installed filters. A potential strategy is to intentionally allocate observation slots that do not synchronize with Torifune's spin period. With this strategy, the planned observations will cover the entire sides of the asteroid. ONC-T will start searching for a potential satellite at T-14 hr.  At T-320 sec, it will continue imaging with different color filters while continuing the search. The final 32 sec before the closest approach will be focused on detailing the asteroid's surface geomorphology at higher resolution by fixing the filter, which is possibly v- or wide (panchromatic)-band. NIRS3 will start its observational sequences at T-1 hr and continue until the spacecraft passes by the asteroid. Within this time range, it will observe compositional variations at different rotational phases, which is about 20\% of the entire surface. TIR will start its sequences at T-30 min, while LIDAR will start at T-4 min. ONC-W1 will be operational only from T-5 min to T-1 min because of observational conflicts. ONC-T and ONC-W1 cannot take images simultaneously. ONC-W1 will join ONC-T for the satellite search and keep tracking the asteroid. 

The flyby sequence planning process accounts for an input from the mission's science team for its preferred viewing geometry and encounter timing. The science team prioritizes the viewing geometry and timing that offer the largest area to be viewed at a visible wavelength, but also some observable nightside areas for infrared flux measurement. The mission developed an assessment tool that accounts for all the employed instruments' observational constraints and requests while also considering the system's constraints, such as thermal and communication conditions. Because the planned flyby speed is 5.25 km s$^{-1}$, remote sensing observations will not be enough to perform global characterizations. The current plan is to fix the spacecraft orientation until T-5 min and change it less than a few degrees toward the target slightly after that. Because all the employed instruments are oriented toward the spacecraft's $-z$ axis, this slight change of the $-z$ axis orientation toward Torifune enables the instruments to capture the asteroid a few seconds before the closest encounter. Without this slight orientation change, NIRS3 will lose Torifune about T-110 sec, and LIDAR will not be able to capture it at all. For LIDAR to receive reflected pulses, the spacecraft needs to be within 25 km (assuming the carbonaceous asteroid case) for its far-system mode \citep{Yamada2017}. It is necessary to find the orientation that allows LIDAR to see Torifune at T-5 sec or later. Thus, this slight orientation change will play a pivotal role in observing Torifune in favored conditions. 

\section{Connections with planetary defense}
\label{Sec:planetary_defense}
Planetary defense is an international effort to assess and mitigate potential threats effectively from various aspects of small bodies that may hit Earth or approach close enough to impact human societies negatively \citep{NAS2022, Chabot2024A}. Hayabusa2\#'s Torifune flyby will be aligned with this effort. One deflection technology is kinetic deflection, which intentionally induces a kinetic impact on a target object to change its {trajectory} away from Earth \citep{NAS2022, NSTC2023, NASAPD2023}. The successful flyby of Torifune will demonstrate Hayabusa2\#'s capabilities to accurately navigate, guide, and control a spacecraft toward a target object. This flyby can also demonstrate how repurposed spacecraft can be placed on intercept trajectories to perform a short-distance flyby of a potentially hazardous object. This section, however, offers perspectives of how Hayabusa2\# is aligned with the fast reconnaissance concept, which was identified as key technology advancement in Planetary Defense \citep{Abell2020,NASAPD2023}. 

NASA/DART was the first full-scale kinetic deflection demonstration mission, which sent a spacecraft to the target asteroid (65803) Didymos, where it successfully collided with its secondary, Dimorphos \citep{Daly2023}. The DART mission successfully demonstrated its sophisticated navigation technique to guide the spacecraft to collide accurately with the target asteroid \citep{Adams2023}. Another goal of this mission was to identify the kinetic deflection efficiency, also known as the momentum transfer enhancement factor, or the so-called $\beta$ value, which defines how much momentum is imparted to the target, from both spacecraft and ejected materials \citep{Rivkin2021, Cheng2023}. One lesson from the DART impact is that determining the $\beta$ value accurately requires knowledge of the target's physical properties prior to impact \citep{Raducan2024, Stickle2025, Hirabayashi2025}. Unfortunately, these quantities were not determined on DART accurately, giving uncertainties in the $\beta$ measurement. The reported $\beta$ value was $3.61^{+0.1}_{-0.25} (1\sigma)$ if Dimorphos's bulk density was assumed to be 2.6 g/cm$^3$. However, if the bulk density was unknown completely, the $\beta$ value would range between 2.2 and 4.9 \citep{Cheng2023}. ESA's Hera visits the Didymos-Dimorphos system and offers critical clues on the system's physical conditions \citep{Michel2022, Michel2025}. 

The lesson from the DART mission is that it is necessary to better characterize a target's key physical properties, such as mass, strength, orbit, and composition, prior to the execution of deflection. One concept to address this issue was to perform rapid investigations to measure (or at least constrain) these properties before the planned deflection \citep{NEOWARP2024}. This process must be performed quickly, particularly when a threat is imminent within a short time frame. The 2023-2032 Decadal Survey led by the National Academies of Sciences, Engineering, and Medicine recommended that the fast reconnaissance concept would be the next planetary defense demonstration mission \citep{NAS2022}, following the NEO Surveyor investigation \citep{Mainzer2023}. The report also identified that this concept would effectively target objects with a diameter ranging between 50 and 100 m \citep{NAS2022}. 

The flyby approach is the quickest for acquiring information concerning the target's physical properties that are relevant for Planetary Defense. Hayabusa2\#'s flyby operation targeting Torifune is relevant to the fast reconnaissance concept, offering an opportunity to demonstrate how the flyby approach can contribute to this concept. Hayabusa2\#'s flyby, however, is not directly aligned with the Decadal Survey's recommended efforts. Because Torifune is a $\sim450$ m diameter asteroid, the accomplishment by Hayabusa2\# does not interfere with future visions and efforts for the concept. Nevertheless, the expected limitation during the flyby is analogous to the fast reconnaissance concept. Therefore, Hayabusa2\# has strong synergy with ongoing and future planetary defense missions such as Hera \citep{Michel2022, Michel2025}, OSIRIS-APEX \citep{DellaGiustina2023}, and RAMSES \citep{Kueppers2023}. 

Another key aspect of Hayabusa2\#'s flyby at Torifune is that the spacecraft was not originally designed and developed for flyby operations, but for rendezvous operations. This will limit operational flexibility, but the mission will need to enhance the spacecraft's capability to maximize engineering and science returns. Furthermore, the mission uses a 10-year-old spacecraft. While the spacecraft is still operational, the system gradually becomes aging and thus degrading. Given those challenges, Hayabusa2\# will demonstrate how employing a used spacecraft can meet the fast reconnaissance concept. Hypothetically, if a spacecraft needs to be developed from scratch and launched from Earth, numerous constraints may hinder rapid operations. However, if there is already an in-flight spacecraft that can reach a target object, using it can be more flexible and less costly, allowing for an earlier mitigation plan.

\section{Conclusion}
Hayabusa2\# is an extended mission of its nominal mission, Hayabusa2, and conducts various engineering and science investigations over its decade-long mission period. This study offered an overview of Hayabusa2\#’s flyby operations at the near-Earth asteroid (162173) Torifune. The mission plans the spacecraft flyby on July 5, 2026. The flyby operation will be a demonstration of engineering technologies for enabling a subkilometer-to-kilometer asteroid flyby encounter and offer stronger insights into the material transport mechanism in the inner solar system, if provided with successful data acquisition from the employed instruments. While the final numbers are to be determined, the flyby speed is planned to be 5.25 km/s, and the closest distance from the asteroid’s center will be between 1 and 10 km or less. During the flyby, the mission will attempt to employ the onboard remote sensing instruments, ONC-T, ONC-W1, TIR, NIRS3, and LIDAR, while not using ONC-W2. Each employed instrument will have measurement goals to achieve the defined science objectives. With the spacecraft’s limited capability to slew, the mission will fix its orientation with slight modification during the flyby. This flyby condition will allow the instruments to capture Torifune a few seconds before the flyby, though not all instruments will have favored observation conditions. Shape reconstruction will also be performed using available data and various approaches to mitigate the expected limited viewing geometry. Establishing a rigorous flyby sequence is a vital step for Hayabusa2\# to obtain meaningful data. Attempting observations under such conditions is synergetic to future Planetary Defense efforts on rapidly determining the physical characteristics of target objects prior to the time of deflection.

\section{Acknowledgments}
This work is supported by JAXA's Hayabusa2\# project. M.H. also acknowledges support from NASA/YORPD (80NSSC25K7696) and the Georgia Institute of Technology.




\end{document}